\documentclass[twocolumn, a4paper]{article}

\date{}

\usepackage{times}
\advance\topmargin-1in\advance\oddsidemargin-1in
\evensidemargin\oddsidemargin

\makeatletter

\def\@normalsize{\@setsize\normalsize{12pt}\xpt\@xpt
\abovedisplayskip 10pt plus2pt minus5pt\belowdisplayskip \abovedisplayskip
\abovedisplayshortskip \z@ plus3pt\belowdisplayshortskip 6pt plus3pt
minus3pt\let\@listi\@listI}

\def\section{\@startsection {section}{1}{\z@}{12pt plus 0pt minus 2pt}
{8pt plus 0pt minus 2pt}{\centering\normalsize\sc
\edef\@svsec{\thesection.\ }}}
\def\thesection{\Roman{section}}

\def\subsection{\@startsection {subsection}{2}{\z@}{10pt plus 0pt minus 2pt}
{6pt plus 0pt minus 2pt}{\normalsize\sl
\edef\@svsec{\thesubsection.\ }}}
\def\thesubsection{\Alph{subsection}}

\long\def\@makecaption#1#2{
\vskip10pt\begin{center} #1 #2 \end{center}\par\vskip 1pt}
\def\fnum@figure{\raggedright{\footnotesize Fig. \thefigure }.%
\footnotesize}
\def\fnum@table{\footnotesize{TABLE} \thetable\\\footnotesize\sc}
\def\thetable{\Roman{table}}

\makeatother

\usepackage{graphicx}
\usepackage{array}
\usepackage{subcaption}
\usepackage{amsmath}
\usepackage{multirow}
\usepackage{setspace}
\usepackage[linesnumbered,lined,commentsnumbered]{algorithm2e}
\usepackage[numbers]{natbib}
\begin{document}

\title{
\Large\textbf{CompRRAE: RRAM-based Convolutional Neural Network Accelerator with Reduced Computations through a Runtime Activation Estimation}\\~\\
\vspace*{-0.8cm}
}	

\author{\normalsize
Xizi Chen, Jingyang Zhu,  Jingbo Jiang and Chi-Ying Tsui \\
\small
Department of Electronic and Computer Engineering, Hong Kong University of Science and Technology, Hong Kong
} 

\maketitle
\thispagestyle{empty}

\small\textbf{Abstract--
Recently Resistive-RAM (RRAM) crossbar has been used in the design of the accelerator of convolutional neural networks (CNNs) to solve the memory wall issue. However, the intensive multiply-accumulate computations (MACs) executed at the crossbars during the inference phase are still the bottleneck for the further improvement of energy efficiency and throughput. In this work, we explore several methods to reduce the computations for the RRAM-based CNN accelerators. First, the output sparsity resulting from the widely employed Rectified Linear Unit is exploited, and a significant portion of computations are bypassed through an early detection of the negative output activations. Second, an adaptive approximation is proposed to terminate the MAC early when the sum of the partial results of the remaining computations is considered to be within a certain range of the intermediate accumulated result and thus has an insignificant contribution to the inference. In order to determine these redundant computations, a novel runtime estimation on the maximum and minimum values of each output activation is developed and used during the MAC operation. Experimental results show that around 70\% of the computations can be reduced during the inference with a negligible accuracy loss smaller than 0.2\%. As a result, the energy efficiency and the throughput are improved by over 2.9 and 2.8 times, respectively, compared with the state-of-the-art RRAM-based accelerators.}

\section{Introduction}

Convolutional neural networks (CNNs) have demonstrated impressive performance in various machine learning tasks such as the visual recognition \cite{AlexKrizhevsky12} and visual tracking \cite{HyeonseobNam15}. At the same time, due to the nature of the convolutional operation, the inference of CNN usually involves intensive computations which are energy consuming and become a big deterrent for deploying CNN in embedded systems. Besides the high computation cost, conventional accelerators also face the memory wall issue where the massive memory accesses for fetching the weights and activations vastly limit the performance. Therefore, it is necessary to deliver a more efficient implementation with fewer computations.

The Rectified Linear Unit (ReLU) \cite{Dahl13} has become the most widely used activation function in neural networks in recent years. Due to the application of ReLU, a high activation sparsity can be achieved during the inference \cite{Albericio16}. Since the negative MAC results will be clamped to zero by ReLU, their actual magnitude values are irrelevant for the cascading layers. Thus, a large portion of computations corresponding to the negative output activations can be bypassed once the sign can be determined early. In addition, the inherent resilience of CNN makes the activation values error-tolerant to some degree, hence making it possible to reduce the computations by approximation without affecting the classification accuracy. To trigger the above computation bypass, we propose a runtime estimation on the maximum and minimum values of each output activation. During each MAC, the contribution of the intermediate accumulated result is evaluated continually against the estimated values of the remaining partial results. Once the contribution of the current accumulated result is considered to be large enough to dictate the final result value, the MAC will be terminated to improve the energy efficiency and the throughput. The proposed methods are implemented in a specialized architecture based on the resistive random access memory (RRAM) crossbar \cite{Wang14} which utilizes the in-situ computation as an approach to address the high power density and the memory wall issue of the conventional CMOS-based design. In summary, the contributions of this work are as follows: 

\begin{itemize}
  \item A runtime estimation on the maximum and minimum values of the output activation is proposed and implemented during each MAC. 
  \item By detecting the negative output activations through the estimation, the corresponding MACs are terminated in advance in the convolutional layers followed by ReLU. According to the experimental results, over 99.98\% of negative outputs are detected and over 71.5\% of their computations are bypassed without inducing accuracy loss.
  \item An adaptive approximation is proposed to bypass the remaining computations during the MAC when the estimated values of the remaining partial results are determined to have a negligible contribution to the inference. 
  \item A dedicated RRAM-based architecture is proposed for implementing the CNNs with reduced computations. A total computation reduction of around 70\% is achieved for the general 16-bit fixed-point implementation, and 40\% reduction is achieved for the 8-bit implementation which demonstrates the effectiveness of the proposed methods under an aggressive quantization scheme. The induced accuracy loss is smaller than 0.2\%. Experimental results show significant improvement in the energy efficiency and throughput.
\end{itemize}

\section{Related Works}

Various techniques have been proposed to reduce the intensive computations in the CNN accelerators. A natural way for reducing the memory footprint and the number of multiplications in the CMOS-based accelerators is to utilize the activation sparsity \cite{VahidehAkhlaghi18,Albericio16,JingyangZhu18}. Since a large portion of input activations are zero, the corresponding multiplications can be bypassed to save energy and time \cite{Albericio16}. A recent work in \cite{VahidehAkhlaghi18} focuses on the layers with only non-negative inputs and exploits the output sparsity by reordering the weights to calculate the sum of the positive products first. Later calculation of the negative products will be terminated as soon as the accumulated result becomes smaller than zero. As a result, 16\% energy saving and 28\% speedup can be achieved. In a more aggressive mode, an empirical value is used to compare with the accumulated result after a specific number of multiplications. If the accumulated result is smaller, the output activation is considered likely to be negative. By bypassing the remaining multiplications, a higher reduction in complexity can be achieved, but a relatively large accuracy loss (3.0\%) is induced. Another work in \cite{JingyangZhu18} exploits the output sparsity by using a low-rank approximation of the weight matrix to predict the output sparsity and disabling the actual computation if the predicted output is negative. In this case, each MAC needs to be broadcast to all the processing elements to improve the throughput. Such methods, however, are not suitable for the RRAM-based architecture. Since the computations are executed at the RRAM crossbars where the weight matrix is programmed into the memristors before the classification, the multiplication-accumulation has to be done in a regular pattern. Therefore, it is difficult to irregularly skip the zero inputs, independently reorder the weights of each kernel, or broadcast different weight matrices to the crossbars at runtime. A way to reduce the computations in the RRAM-based architecture is to structurally compress the weights through training and then exploit the weight sparsity \cite{ReCom18}. However, to the best of our knowledge, the output sparsity hasn't been efficiently exploited in the RRAM-based architecture. Due to the resilience of CNN, reducing the quantization bit-width of the weights and activations is another method for reducing computations \cite{Moons16}. A dynamic quantization scheme is proposed in \cite{XiziChen18} to change the bit-width of the weights when multiplying with the different bit of the activations to reduce the computations for the RRAM-based MAC.

\section{Preliminaries}

\subsection{Convolutional Neural Networks}

A convolutional neural network (CNN) \cite{AlexKrizhevsky12} is a machine learning model inspired by the structure of the human brain. It is usually comprised of a series of cascading layers, including the convolutional (CONV) layers, pooling layers, and fully-connected (FC) layers. The CONV and FC layers consist of neurons to extract the features of the image. Inside each layer, the input activations from the previous layer are firstly multiplied with the corresponding weights in the current layer, and then accumulated to generate the output activations for the next layer. The computation of the CONV layer is shown in Fig.\ref{fig: crossbar}(a) and can be expressed as follows:

\begin{small}
\begin{equation}\label{eq_conv}
\setlength{\abovedisplayskip}{3pt}
\setlength{\belowdisplayskip}{3pt}
a_{out}(x,y,z)=
f(\sum^{c-1}_{l=0}\sum^{h-1}_{m=0}\sum^{w-1}_{n=0}
a_{in}(x+m,y+n,l) \times K_z(m,n,l))
\end{equation}
\end{small}
\vspace{-5pt}
 
where \begin{small}$a_{out}$\end{small} and \begin{small}$a_{in}$\end{small} represent the output and the input activations, respectively; \begin{small}$K_z$\end{small} represents the \begin{small}$z^{th}$\end{small} kernel; \begin{small}$h$\end{small}, \begin{small}$w$\end{small} and \begin{small}$c$\end{small} represent the height, width, and depth of the kernel; \begin{small}$(x,y,z)$\end{small}, \begin{small}$(x+m,y+n,l)$\end{small} and \begin{small}$(m,n,l)$\end{small} represent the positions of the activations and weights in height, width, and depth. \begin{small}$f$\end{small} is a non-linear activation function to avoid overfitting. The most commonly used activation function is ReLU given by: 

\begin{small}
\vspace{-4pt}
\begin{equation}\label{eq_relu}
\setlength{\abovedisplayskip}{3pt}
\setlength{\belowdisplayskip}{3pt}
f(x)= max(0,x)
\end{equation}
\end{small}
\vspace{-4pt}

where $x$ is the input to the function. FC layers are similar to the CONV layers, but have much fewer computations which can be simplified to a single vector-matrix multiplication. Pooling layers usually follow the CONV layers for down-sampling. In this work we will focus on the CONV layers since they account for most of the computations in CNN.

\subsection{RRAM Crossbar and In-Situ Computation}

\begin{figure}[tb]
\centering
	\includegraphics[width=0.46\textwidth]{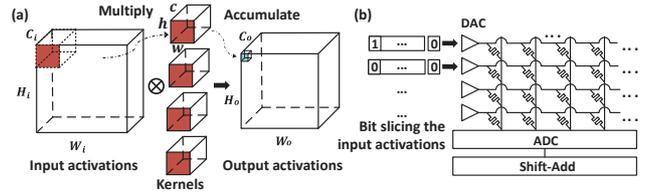} 	
	\caption{\small 	
	(a) Computation of the CONV Layer; (b) In-Situ Computation based on the RRAM Crossbar.
	}
	\label{fig: crossbar}
\end{figure}

The RRAM crossbar has aroused great research interest due to its high-density, non-volatility, and the potential for parallel in-situ analog computation \cite{Shafiee16,Wang14,Feinberg18}. While the CMOS-based accelerators face the difficulty of scaling down and the issue of memory wall, the RRAM-based computation provides a promising approach to achieve substantial improvement in the energy efficiency and throughput. For instance, the RRAM-based accelerator in \cite{Wang14} demonstrates 22 times energy saving compared with the CMOS-based counterpart. A hierarchical RRAM-based architecture proposed in \cite{Shafiee16} improves the energy efficiency and throughput by 5.5 and 14.8 times, respectively, compared with the state-of-the-art CMOS-based DaDianNao architecture \cite{DaDianNao14}. The RRAM crossbar for the vector-matrix multiplication is shown in Fig.\ref{fig: crossbar}(b). The elements of the weight matrix are stored as the conductance values of the memristors at the crosspoints connecting the horizontal wordlines and the vertical bitlines. When the computation starts, the input activation vector is applied on the wordlines as voltages. The current flowing through each memristor is equal to the product of the memristor conductance and the wordline voltage. Currents on the same bitline will be accumulated and output as the computation result. Since the computation is done in analog, digital-to-analog convertors (DACs) are needed at the wordlines to convert the input activations to voltages, and analog-to-digital convertors (ADCs) are needed at the bitlines to convert the results back to digital values. These interfacing circuits are the most energy consuming part during the computation \cite{Shafiee16,XiziChen18}. Since hundreds of products are accumulated vertically, the resolution requirement of ADC can easily go beyond the acceptable range and induce a huge energy overhead. As a common solution, the multi-bit multiplication, e.g. 16-bit multiplication, is broken into a series of low bit-width multiplications to limit the ADC resolution within a reasonable range \cite{Feinberg18,Shafiee16,XiziChen18}. For example, for a crossbar with 128 wordlines, the resolution of each crosspoint should be no more than 2-bit to keep the ADC resolution less than 10-bit. Thus, each weight takes multiple memristors to store. At the same time, since the multi-bit DAC is expensive to implement and hundreds of DAC operations are needed for one MAC, it is more efficient and preferable to use the single-bit DAC to minimize the overhead \cite{Shafiee16,XiziChen18}. Thus, each bit of the input activation is sent into the crossbar sequentially to finish the MAC. The whole MAC operation will take multiple iterations. At each iteration, a partial result corresponding to the current 1-bit input vector is generated, and then accumulated with the existing results of previous iterations. This bit-level slicing of activations creates a special scheduling for MAC and we will utilize this characteristic to reduce the computations.

\section{RRAM-based Computation Reduced Accelerator Design}

\subsection{Algorithms for Computation Reduction}

\begin{figure}[tb]
\centering
	\includegraphics[width=0.485\textwidth]{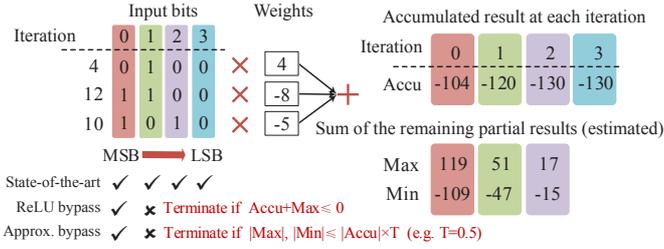} 	
	\caption{\small{Computation Reduction in the RRAM-based Accelerator}}
	\label{fig: Algorithms}
\end{figure}

In the CMOS-based accelerators, the MAC is usually done by accumulating the corresponding activation-weight products. However, since the activations are sliced into bit-level in the RRAM-based architecture, only a partial result is obtained at each iteration by accumulating the input bit-weight products. As the MAC sequentially proceeds from the most-significant bit (MSB) to the least-significant bit (LSB) of the input activations, the generated partial results also become less and less significant. Each new generated partial result will be accumulated with the sum of the partial results of previous iterations. An example is given in Fig.\ref{fig: Algorithms} to illustrate this computation process, where the inner product of the activations \begin{small}$[4,12,10]$\end{small} and the weights \begin{small}$[4,-8,-5]$\end{small} is computed in 4 iterations. The intermediate accumulated result is updated at each iteration, represented as \textbf{Accu} \begin{small}$[-104,-120,-130,-130]$\end{small} in Fig.\ref{fig: Algorithms}. Such bit-level processing makes it possible to terminate the MAC in advance once the remaining iterations are considered to be redundant based on the possible values of their partial results estimated beforehand. Specifically, two schemes are exploited to identify the redundant iterations, as described below. 

\begin{figure}[tb]
\centering
	\includegraphics[width=0.38\textwidth]{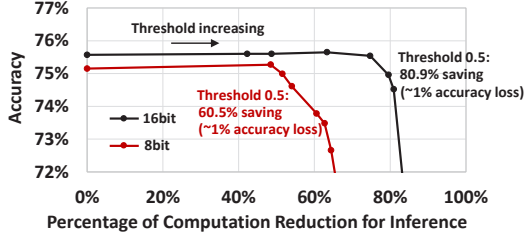} 	
	\caption{\small Ideal Performance of the Adaptive Approximation for the 16-bit and 8-bit Implementations}
	\label{fig: Ideal Performance of Adaptive Approximation}
\end{figure}

\textit{1) ReLU-based Computation Bypass:} 
For an early identification of the negative outputs followed by ReLU, the maximum value of the sum of the partial results for the remaining iterations is estimated beforehand and represented as \textbf{Max} in Fig.\ref{fig: Algorithms}. (The method for estimating \textbf{Max} will be elaborated later.) If at any iteration, the sum of \textbf{Accu} and \textbf{Max} is not larger than zero (\begin{small}$Accu+Max\leq0$\end{small}), the final output is considered to be non-positive and hence the MAC can be terminated in advance. Otherwise, the MAC will continue. For instance, the MAC in Fig.\ref{fig: Algorithms} will be terminated after the second iteration since \begin{small}$-120+51\leq0$\end{small}. It will be shown in the experimental results that the negative outputs of the CONV layers account for 57.5\% of the total computations in the CifarQuick Model \cite{YangqingJia14} on Cifar-10 \cite{articleCifar}, and 71.5\% of their computations can be bypassed based on the estimation.

\textit{2) Adaptive Approximation:} 
For the activations not supported by the ReLU-based bypass, adaptive approximation is proposed for the early termination of MAC. In general, the resilience of network allows the activations to deviate from their actual values within a certain range without affecting the classification result. Based on this, an adaptive approximation scheme is proposed as shown in Fig.\ref{fig: Algorithms}. If the magnitude of \textbf{Max} and \textbf{Min} (Min: minimum value of the sum of the remaining partial results) is not larger than a certain threshold (\textbf{T}) of the magnitude of \textbf{Accu} (\begin{small}$|Max|,|Min|\leq|Accu|\times T$\end{small}), which reflects the allowable deviation from the actual result, the MAC can be terminated. For instance, the last two iterations can be bypassed if \textbf{T} is set as 0.5 in Fig.\ref{fig: Algorithms}. The allowable deviation for triggering the bypass varies adaptively with \begin{small}$|Accu|$\end{small}. The larger the \begin{small}$|Accu|$\end{small}, the larger the allowable deviation is. \textbf{T} is an empirical tunable parameter to balance the accuracy and the complexity saving. A larger amount of computation reduction can be achieved by increasing \textbf{T}, but the accuracy loss will also increase at the same time. To obtain the ideal performance, i.e. the upper bound of complexity saving of the adaptive approximation, we assume we know the exact value of the output activation beforehand and so the actual partial result at each iteration is used instead of the estimated values. Based on this, we can exactly know which iterations will not be needed and the ideal maximum amount of bypass can be obtained. This ideal performance and the upper bound of saving at different \textbf{T} values for the general 16-bit fixed-point implementation of CifarQuick is shown in Fig.\ref{fig: Ideal Performance of Adaptive Approximation}. A sweet spot is observed where 80.9\% computations can be reduced with an accuracy loss of 1\% for an optimum threshold value. Similarly, for the 8-bit fixed-point implementation, ideally a computation reduction of 60.5\% can be achieved by the adaptive approximation at the same threshold as shown in Fig.\ref{fig: Ideal Performance of Adaptive Approximation}. The savings shown in Fig.\ref{fig: Ideal Performance of Adaptive Approximation} assume a perfect knowledge of the partial result at each iteration. However in real situation, the actual partial results will not be known beforehand. Therefore we propose a method to get an accurate estimation on the maximum and minimum values of the partial result at each iteration.

\subsection{Making a Runtime Estimation on the Output Activation}

Considering the MAC operation for an output activation, the worst-case maximum value of each partial result can be estimated by assuming all the input bit-weight products to be accumulated are as large as possible. In this case, for the layers with both positive and negative inputs, the maximum value of the partial result is equal to \begin{small}$\sum|w| \times 2^i$\end{small} where \begin{small}$w$\end{small} represents the weight in the kernel and \begin{small}$i$\end{small} is the position of the input bit from bit-width\begin{small}$-$\end{small}1 to 0. Similarly, the worst-case minimum value of each partial result is equal to \begin{small}$-\sum|w|\times 2^i$\end{small}. For the hidden layers following ReLU, since there is no negative input, the maximum and minimum values of the partial result become\begin{small} $\sum w_+ \times 2^i$\end{small} and \begin{small}$\sum w_- \times 2^i$\end{small}, where \begin{small}$w_+$\end{small} and \begin{small}$w_-$\end{small} represent the positive and the negative weights, respectively. The computation bypass based on this loose bound of estimation ensures no accuracy loss, but the amount of the complexity reduction is small since the worst-case estimated values usually have much larger magnitude than the actual result due to several reasons. First, since the multi-bit input activation has been broken into multiple iterations, it is highly likely to have some input bits equal to zero even when the activation is positive. Thus, a considerable portion of input bit-weight products are actually zero. Moreover, in the worst case, all the input bit-weight products to be accumulated are assumed to have the same sign. However, in practice each non-zero product can either have positive impact or negative impact on the partial result. 

\begin{figure}[tb]
\centering
	\includegraphics[width=0.48\textwidth]{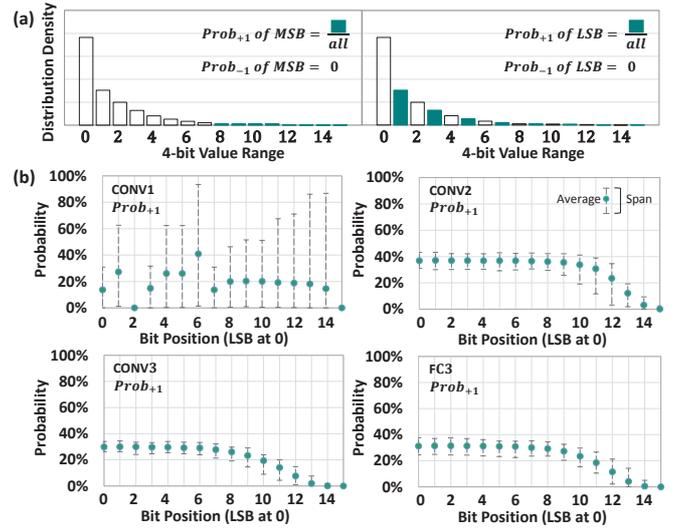} 	
	\caption{\small (a) An Example for Extracting the Probabilities from the Distribution of the 4-bit Input Activations; (b) Average Probabilities and the Spans Extracted for the 16-bit implementation of CifarQuick.}
	\label{fig: Probabilities}
\end{figure}

To have a more practical estimation, a tighter bound based on the actual input activation statistics is proposed. Before the classification, the empirical activation statistics of each layer are obtained from the training images, and the probabilities of the input bit at each iteration to be \begin{small}$+1$\end{small} and \begin{small}$-1$\end{small} are calculated accordingly. As an example, Fig.\ref{fig: Probabilities}(a) illustrates how to extract the corresponding probabilities for the MSB and LSB from the distribution of the 4-bit input activations in a CONV layer of CifarQuick. Specifically, the probability of MSB to be \begin{small}$+1$\end{small} is equal to the occurrence probability of the activation that has a value larger than 8. It can be easily extended to the implementations with different bit-widths. For a larger bit-width such as 16-bit, the activations will be partitioned into smaller bins. The average value and the span of the probability of each input bit to be \begin{small}$+1$\end{small} extracted for the 16-bit implementation of CifarQuick are shown in Fig.\ref{fig: Probabilities}(b). For the hidden layers, the probability of the MSB to be \begin{small}$+1$\end{small} is small since most of the activations have small values. For the LSB, the probability is within the range from 25\% to 40\% since a large portion of input activations are zero. The probability of having a \begin{small}$-1$\end{small} input is zero for the layers following ReLU. CONV1 is different from others due to the mean subtraction for image pre-processing. It is worth noting that the activation distribution normally doesn't change much for different images in the dataset, and thus the empirical probabilities can be generally utilized for the estimation. The maximum and minimum values of the partial result estimated at a specific iteration $i$ ($i$=0 for LSB) are given by:

\begin{small}
\begin{equation}\label{eq_conv}
\begin{split}
&max = (max_{+} + max_{-})\times 2^i,    min = (min_{+} + min_{-})\times 2^i\\
&max_{+} = \sum{w_+}\times Prob_{+1, max} + \sum{|w_-|}\times Prob_{-1, max}\\
&max_{-} = \sum{-w_+}\times Prob_{-1, min} + \sum{w_-}\times Prob_{+1, min} \\
&min_{+} = \sum{w_+}\times Prob_{+1, min} + \sum{|w_-|}\times Prob_{-1, min} \\
&min_{-} = \sum{-w_+}\times Prob_{-1, max} + \sum{w_-}\times Prob_{+1, max} \\
\end{split}
\end{equation}
\end{small}

where \begin{small}$max$\end{small} and \begin{small}$min$\end{small} represent the estimated maximum and minimum values of the partial result, respectively. To estimate \begin{small}$max$\end{small}, the input bit-weight products with different signs need to be separately considered. We first consider the case where the input bits and the corresponding weights have the same signs to estimate the sum of the positive products (\begin{small}$max_{+}$\end{small}). \begin{small}$\sum w_+$\end{small} and \begin{small}$\sum w_-$\end{small} represent the sums of the positive weights and the negative weights in the kernel, respectively. For a better accuracy, a conservative bound should be used for the estimation, and thus we use \begin{small}$Prob_{+1,max}$\end{small} and \begin{small}$Prob_{-1,max}$\end{small} to estimate \begin{small}$max_{+}$\end{small} where \begin{small}$Prob_{+1,max}$\end{small} and \begin{small}$Prob_{-1,max}$\end{small} represent the maximum probabilities of the input bit to be \begin{small}$+1$\end{small} and \begin{small}$-1$\end{small}, respectively. Also we need to estimate the sum of the negative products (\begin{small}$max_{-}$\end{small}) when the input bits and weights are of opposite signs. Again to have a conservative bound we use \begin{small}$Prob_{+1,min}$\end{small} and \begin{small}$Prob_{-1,min}$\end{small} which represent the minimum probabilities of the input bit to be \begin{small}$+1$\end{small} and \begin{small}$-1$\end{small}, respectively. The minimum value of the partial result (\begin{small}$min$\end{small}) can be estimated in a similar way. This statistics-based estimation is more precise than the worst-case estimation.

\subsection{Hardware Architecture}

\begin{figure}[tb]
\centering
	\includegraphics[width=0.48\textwidth]{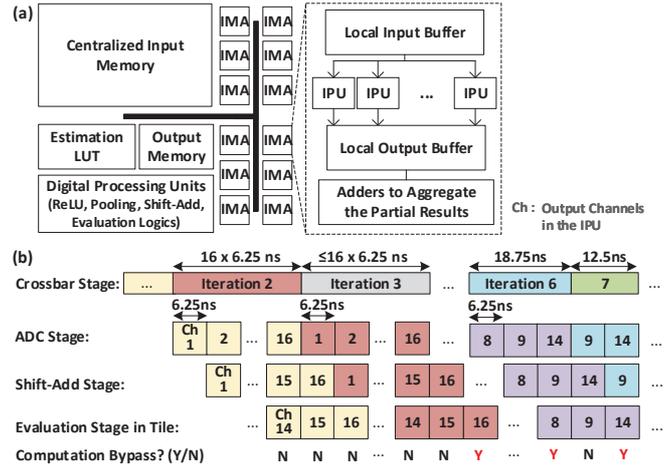} 	
	\caption{\small (a) Hardware Architecture inside a Tile; (b) The Pipeline of CompRRAE.}
	\label{fig: Architecture}
\end{figure}

The analog computation is done inside the in-situ processing units (IPUs) similar as Fig.\ref{fig: crossbar}(b). Each IPU contains a group of 1-bit DACs at the input of the wordlines, a pair of differential RRAM crossbars to store the positive and negative weights, respectively, the sample-hold units to hold the bitline currents, a single ADC which is time-shared by the bitlines, and a shift-add unit to aggregate the partial results after ADC for the MAC operation. In order to make a comparison with the state-of-the-art RRAM-based accelerator, we adopt a hierarchical architecture similar to ISAAC presented in \cite{Shafiee16} as the baseline, and compare the proposed CompRRAE with ISAAC in terms of the energy efficiency, throughput, and area cost. Similar to ISAAC, each accelerator contains multiple tiles connected with a concentrated-mesh at the top level. The architecture inside a tile is shown in Fig.\ref{fig: Architecture}(a). Each tile contains multiple in-situ multiply-accumulate modules (IMAs) sharing the same centralized memories and digital processing units which are used to execute digital operations such as shift-add, ReLU and pooling. Inside the tile, the IMAs are connected through a shared bus. Inside each IMA, there are multiple IPUs sharing a local input buffer and a local output buffer which hold the input and output activations, respectively, during the MAC. 

In order to implement the proposed runtime estimation, first, the probabilities are extracted offline. Based on Eq.(3), the estimated maximum and minimum values of the partial result at each iteration are computed independently for each output channel. Then, the corresponding estimated partial results are summed up to get the \textbf{Max} and \textbf{Min} for each iteration. Same as the example shown in Fig.\ref{fig: Algorithms}, there are totally N-1 \textbf{Max} and \textbf{Min} values for each output channel, where N is the number of iterations, i.e. the bit-width of the activation\begin{small}$-$\end{small}1. This process is done offline and the estimated \textbf{Max} and \textbf{Min} values are stored in a look-up table (LUT) in the tile. During runtime, at each iteration, the estimated value of the output activation is the sum of the actual accumulated result and the estimated value for the remaining iterations read from the LUT. The evaluation logics to compute the formulas in Fig.\ref{fig: Algorithms} for deciding whether to skip the remaining iterations include an adder for the ReLU-based bypass and a multiplier and comparator for the approximation-based bypass.  The size of the tile is normally large enough for mapping a complete kernel of the CONV layers. However, each kernel may occupy multiple IMAs and thus the local results in the IMAs are required to be sent out through the shared bus and aggregated in the tile. To minimize the overhead of data transfer, each kernel is preferred to fully occupy the IPUs in one IMA first before occupying others 
when mapping the network, and the calculated results are firstly aggregated locally inside the IMAs before sent out through the shared bus. 

We use the same configuration as that in ISAAC where each memristor has 2-bit precision in the 128$\times$128 RRAM crossbar. Since each 16-bit weight takes 8 memristors to store, there are 16 output channels mapped to one IPU. The pipeline schedule of CompRRAE is shown in Fig.\ref{fig: Architecture}(b). After finishing each crossbar computation, the bitline results are firstly latched in the sample-hold circuits. In the next stage, a 1.28GHz ADC sequentially converts the 8 bitline currents for an output activation in 6.25ns in the IPU. The 8 bitline results are processed by the shift-add to generate a local partial result in the IPU in the next 6.25ns. Then, in the next stage, the local partial results of different IMAs are aggregated in the tile and update the intermediate accumulated result of the MAC. This result together with the estimated values stored in the LUT will be sent to the evaluation logics to decide whether the computation can be terminated earlier based on the algorithms presented in Section IV, and the control signals will be generated and sent back to the IMAs. All these operations will be finished in the 6.25ns time frame. At the beginning of the MAC operation, i.e. the first iteration, it takes 16$\times$6.25ns to finish the iteration for the 16 output activations in the IPU. As the computation goes on, some of the output activations may be bypassed and the time for each iteration in the IPU may become shorter. After all the IPUs finish the computation, the results will be sent to the next layer, and the next MAC operation of the current layer will start. Since the iterations may take less time in CompRRAE due to the computation bypass, the bandwidth of the input memories to provide the necessary input activations to the IPUs have to be increased. If we need to maintain the input memory bandwidth the same as that in ISAAC, the number of IPUs in each tile needs to be reduced to make sure the memory bandwidth can support the IPUs which are now running faster. At the same time, since less IPUs are used in each tile, more tiles are needed for mapping the same network and this will cause area overhead.

Compared with ISAAC, the energy overhead mainly comes from the evaluation logics, the additional data transfers through the shared bus, and the extra memory accesses of the LUT and the centralized output buffer. Extra area overhead will be required for the evaluation logics and the LUT. The detailed analysis will be discussed in the next section.

\section{Experimental Result}

\subsection{Models of Energy, Area, and Throughput}

The operation parameters and the corresponding energy and area data for the major components of CompRRAE are summarized in Table \ref{tab: power and area}. All the memories and the shared buses are modeled at 32nm in CACTI6.5 \cite{Muralimanohar07}. The centralized input memory is implemented using eDRAM. The local buffers, the centralized output memory and the LUT are implemented using SRAM. The conductance range and the area of RRAM are taken from the Stanford-PKU RRAM model \cite{ZizhenJiang14}, and the corresponding power is simulated using a device-level simulator implemented in C++. The parameters of the 1-bit DAC are obtained through a real design implemented in Cadence at TSMC 65nm and scaled shown to 32nm process. Same as ISAAC, an 8-bit SAR ADC is adopted, and the power and area are taken from \cite{Kull13}. The evaluation logics are designed and implemented in Verilog and synthesized using TSMC 65nm. The power and area are obtained and scaled down to 32nm process. The parameters of other digital processing units such as the shift-add and the sample-hold are adapted from ISAAC \cite{Shafiee16}. The time for the IPUs to finish the MAC operation for each layer is used to model the execution time, and a simulation-based throughput model is built in SystemC. We also calculate the corresponding energy, throughput, and area of ISAAC as a baseline to compare with.

\begin{table}[tb]
\renewcommand{\arraystretch}{1.2}
\centering
\caption{\small Power and Area Estimation}
\begin{scriptsize}
\begin{tabular}{|l|l|l|l|}
\hline    
\multicolumn{4}{|c|}{\textbf{Centralized Memories and Buses}} \\
\hline
\textbf{Component} & \textbf{Spec} & \textbf{Energy}($nJ$) 	& \textbf{Area}($um^2$)\\	
\hline
Input Memory			& size: 64KB							& 0.0188								& \multirow{2}{*}{46000}\\
(eDRAM)					& bus width: 256bit				& (0.38mW leakage)				& \\
\hline
Output Memory		& size: 1KB							& 0.0008								& \multirow{2}{*}{3900}\\
(SRAM)					& bus width: 128bit				&(0.13mW leakage)				& \\
\hline
Estimation LUT			& size: 5KB			 				& 0.0035								& \multirow{2}{*}{9600}\\
(SRAM)					& bus width: 160bit				&(0.002mW leakage)			& \\
\hline
Bus of						& num: 256    						& \multirow{2}{*}{0.0042	}	& \multirow{2}{*}{80000}\\
Input Path				& delay: 0.44ns 					&											&\\
\hline
Bus of						& num: 128							& \multirow{2}{*}{0.0020	}	& \multirow{2}{*}{39100}\\	    
Output Path				& delay: 0.43ns 					&											&\\
\hline
\hline
\multicolumn{4}{|c|}{\textbf{Local Memories (Shared among 8 IPUs)}} \\
\hline
\textbf{Component} & \textbf{Spec} & \textbf{R/W Energy}($nJ$) & \textbf{Area}($um^2$)\\	
\hline
\multirow{2}{*}{Input Buffer}		& size: 2KB					& 0.0019								& \multirow{2}{*}{7400}\\
    													& bus width: 256bit		& (0.42mW leakage)				& \\
\hline
\multirow{2}{*}{Output Buffer}	& size: 256B					&0.0005								& \multirow{2}{*}{2600}\\
    													& bus width: 128bit		&(0.05mW leakage)				& \\   
\hline
\hline
\multicolumn{4}{|c|}{\textbf{IPU Parameters at 1.28GHz (80 MACs per Tile)}} \\
\hline
\textbf{Component} & \textbf{Spec} & \textbf{Power}($mW$) & \textbf{Area}($um^2$)\\	
\hline
\multirow{2}{*}{DAC} 	& resolution: 1 bit 		& \multirow{2}{*}{0.25}  	& \multirow{2}{*}{668}\\
          							& num: 256 			 		& 										& \\
\hline
\multirow{2}{*}{ADC} 	& resolution: 8 bit 		& \multirow{2}{*}{3.1} 		& \multirow{2}{*}{1500}\\
          							& num: 1  						&  									& \\
\hline
Memristor 	 				& resolution: 2 bit 		& Cifar-10: 1.5 					& \multirow{2}{*}{264}\\
Crossbar    					& num: 2 				 		& MNIST:  0.7	 				& \\
\hline
Sample-Hold				& num: 128					& 0.001							& 5 \\
\hline
Shift-Add      					& num: 1 						& 0.05 								& 60 \\
\hline
\hline
\multicolumn{4}{|c|}{\textbf{Other Tile Parameters at 1.28GHz}} \\
\hline
\textbf{Component} & \textbf{Spec} & \textbf{Power}($mW$) & \textbf{Area}($um^2$)\\	
\hline
Evaluation Logic		& num: 8							& 0.79								& 320 	\\
\hline
Shift-Add					& num: 8							& 0.4								& 480 	\\
\hline
\end{tabular}
\label{tab: power and area}
\end{scriptsize}
\end{table}

\subsection{Benchmarks}

We use two benchmarks to compare CompRRAE with ISAAC. The first benchmark is the LeNet-5 \cite{Lecun98} which has two CONV layers and two FC layers trained on the handwritten digit dataset MNIST. The second benchmark is the medium-sized CifarQuick model \cite{YangqingJia14} with three CONV layers and two FC layers trained on the color image dataset Cifar-10 \cite{articleCifar}. The proposed schemes are firstly tested for a 16-bit quantization for comparison with ISAAC, and then tested for an 8-bit quantization to demonstrate the effectiveness of CompRRAE under an aggressive quantization scheme. The accuracy of the fixed-point implementations are summarized in Table \ref{tab: Accuracy of the Fixed-Point Implementations}.

\begin{table}[tb]
\renewcommand{\arraystretch}{1.2}
\centering 
\caption{\small Accuracy of the Fixed-Point Implementations}
\begin{scriptsize}
\begin{tabular}{|>{\centering\arraybackslash} m{2.2cm}|>{\centering\arraybackslash} m{2.5cm}|>{\centering\arraybackslash}m{2.5cm}|}
\hline
\textbf{Benchmarks} 		& \textbf{16-bit Representation} 		& \textbf{{\centering 8-bit Representation}}	\\
\hline
CifarQuick			& 75.57\%						& 	75.15\%  \\
\hline
LeNet-5				& 99.13\%						& 99.09\%  \\
\hline
\end{tabular}
\label{tab: Accuracy of the Fixed-Point Implementations}
\end{scriptsize}
\end{table}

\begin{table}[tb]
\renewcommand{\arraystretch}{1.2}
\centering
\caption{\small Results of the ReLU-based Computation Reduction in CifarQuick}
\begin{scriptsize}
\begin{tabular}{|>{\centering\arraybackslash} m{2.8cm}|>{\centering\arraybackslash} m{2.2cm}|>{\centering\arraybackslash}m{2.2cm}|}
\hline
\textbf{Computation Reduction} 		& \textbf{16-bit Implementation} 					& \textbf{8-bit Implementation}	\\
\hline
For the Negative Outputs				& 	\multirow{2}{*}{71.5\%}	& 	\multirow{2}{*}{44.1\%}	\\
followed by ReLU							&		& \\
\hline
For a Complete Inference				& 40.2\%	& 23.8\%	\\
\hline
\end{tabular}
\label{tab: ReLU-based computation reduction}
\end{scriptsize}
\end{table}

\begin{table*}[tb]
\renewcommand{\arraystretch}{1.2}
\centering
\caption{\small The Overall Computation Reduction, Energy Efficiency, Throughput, and Area Efficiency}
\begin{scriptsize}
\begin{tabular}{|c|cc|cc|cc|cc|} 
\hline
\multirow{2}{*}{\textbf{Performance}} 	& \multicolumn{2}{c|}{\textbf{16-bit CifarQuick}}  	& \multicolumn{2}{c|}{\textbf{8-bit CifarQuick}}	
																& \multicolumn{2}{c|}{\textbf{16-bit LeNet-5}}  		& \multicolumn{2}{c|}{\textbf{8-bit LeNet-5}}	\\
\cline{2-9}
																& \multicolumn{1}{c|}{\textbf{Baseline}} & \multicolumn{1}{c|}{\textbf{CompRRAE}} 
																& \multicolumn{1}{c|}{\textbf{Baseline}} & \multicolumn{1}{c|}{\textbf{CompRRAE}} 
																& \multicolumn{1}{c|}{\textbf{Baseline}} & \multicolumn{1}{c|}{\textbf{CompRRAE}}
																& \multicolumn{1}{c|}{\textbf{Baseline}} & \multicolumn{1}{c|}{\textbf{CompRRAE}}  \\
\hline
Accuracy							& \multicolumn{1}{c|}{75.57\%}	& \multicolumn{1}{c|}{75.44\%} 	
										& \multicolumn{1}{c|}{75.15\%}	& \multicolumn{1}{c|}{75.00\%}		
										& \multicolumn{1}{c|}{99.13\%}	& \multicolumn{1}{c|}{98.97\%}
										& \multicolumn{1}{c|}{99.09\%}	& \multicolumn{1}{c|}{98.90\%} \\
\hline
Computation Reduction for a Complete Inference	& \multicolumn{1}{c|}{-}	& \multicolumn{1}{c|}{69.4\%}		
										& \multicolumn{1}{c|}{-}	& \multicolumn{1}{c|}{39.1\%}	
										& \multicolumn{1}{c|}{-}	& \multicolumn{1}{c|}{78.5\%}
										& \multicolumn{1}{c|}{-}	& \multicolumn{1}{c|}{45.4\%} \\
\hline
Energy Consumption ($mJ/frame$)& \multicolumn{1}{c|}{5.80e-2}	& \multicolumn{1}{c|}{2.00e-2}		
										& \multicolumn{1}{c|}{1.41e-2}	& \multicolumn{1}{c|}{1.00e-2} 	
										& \multicolumn{1}{c|}{1.71e-2}	& \multicolumn{1}{c|}{5.65e-3}
										& \multicolumn{1}{c|}{4.31e-3}	& \multicolumn{1}{c|}{2.61e-3} \\
\hline
Energy Efficiency ($frames/J$)& \multicolumn{1}{c|}{1.72e+4}	& \multicolumn{1}{c|}{5.00e+4}		
										& \multicolumn{1}{c|}{7.10e+4}	& \multicolumn{1}{c|}{1.00e+5} 	
										& \multicolumn{1}{c|}{5.83e+4}	& \multicolumn{1}{c|}{1.77e+5}
										& \multicolumn{1}{c|}{2.32e+5}	& \multicolumn{1}{c|}{3.83e+5} \\
\hline
Throughput ($frames/s$)& \multicolumn{1}{c|}{603.0}	& \multicolumn{1}{c|}{1715.6}		
										& \multicolumn{1}{c|}{1205.2}	& \multicolumn{1}{c|}{1978.3} 	
										& \multicolumn{1}{c|}{1082.0}	& \multicolumn{1}{c|}{4897.1}
										& \multicolumn{1}{c|}{2158.0}	& \multicolumn{1}{c|}{4163.1} \\
\hline
Area ($mm^2$)& \multicolumn{1}{c|}{0.5125}	& \multicolumn{1}{c|}{0.5816}		
										& \multicolumn{1}{c|}{0.3022}	& \multicolumn{1}{c|}{0.3468} 	
										& \multicolumn{1}{c|}{1.5375}	& \multicolumn{1}{c|}{1.7252}
										& \multicolumn{1}{c|}{0.8204}	& \multicolumn{1}{c|}{0.9243} \\
\hline
Area Efficiency ($frames/s/mm^2$)& \multicolumn{1}{c|}{1176.8}	& \multicolumn{1}{c|}{2949.8}		
										& \multicolumn{1}{c|}{3988.1}	& \multicolumn{1}{c|}{5704.4} 	
										& \multicolumn{1}{c|}{703.7}	& \multicolumn{1}{c|}{2838.5}
										& \multicolumn{1}{c|}{2630.4}	& \multicolumn{1}{c|}{4504.0} \\
\hline
\end{tabular}
\label{tab: Overall Performance}
\end{scriptsize}
\end{table*}

\subsection{Results of the ReLU-based Computation Reduction}

The negative output activations in the CONV layers of CifarQuick account for 57.5\% of the total computations during the inference. On average, over 99.9\% of the negative outputs are detected based on the runtime estimation and over 71.5\% of the computations corresponding to these negative outputs are reduced for the 16-bit implementation. For the 8-bit implementation, over 98.3\% of the negative outputs are detected and 44.1\% of their computations are reduced. Thus, the overall ReLU-based computation reduction for a complete inference are 40.2\% and 23.8\% for the 16-bit and 8-bit implementations, respectively. No accuracy has been compromised for both implementations. Since the CONV layers of LeNet-5 are not followed by ReLU, the computation will only be reduced by the adaptive approximation. The performance of the ReLU-based computation bypass in CifarQuick is summarized in Table \ref{tab: ReLU-based computation reduction}.

\begin{figure}[tb]
\centering
	\includegraphics[width=0.485\textwidth]{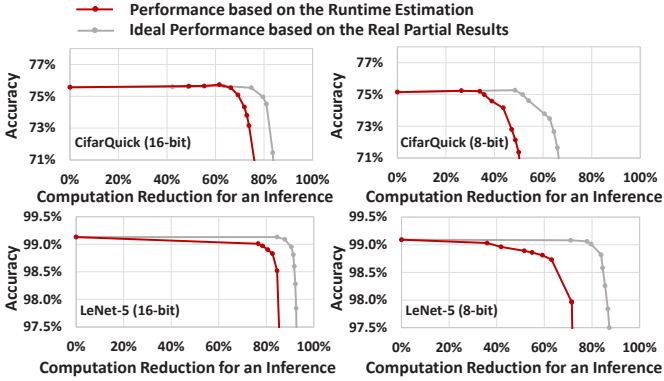} 	
	\caption{{\small Results of the Adaptive Approximation}}
	\label{fig: Approximation bypass result}
\end{figure}

\subsection{Results of the Adaptive Approximation}

The results of the computation bypass based on the adaptive approximation are shown in Fig.\ref{fig: Approximation bypass result}. To maximize the amount of computation reduction while maintaining a high accuracy, the optimal threshold for the approximation is empirically found as 0.8 for the 16-bit implementation of CifarQuick, where 67.4\% of computations can be reduced for the inference with an accuracy loss as small as 0.13\%. For the 8-bit implementation of CifarQuick, 35.8\% computation reduction is achieved at the same threshold for the inference with only 0.16\% accuracy loss. Similar trend has been observed in LeNet-5, where 78.5\% and 45.4\% computation reductions are obtained for the 16-bit and 8-bit implementations, respectively. The accuracy loss is smaller than 0.19\%.

\subsection{Overall Performance and Overhead Analysis}

The overall computation reduction, energy efficiency improvement, and throughput improvement after combining the two proposed schemes are summarized in Table \ref{tab: Overall Performance}. The overall computation reduction achieved for the 16-bit implementation of CifarQuick is 69.4\% with 0.13\% induced accuracy loss. Therefore, the energy efficiency and throughput are improved by 2.9 times and 2.8 times, respectively. The energy overhead caused by the runtime estimation (i.e. the evaluation logics, the extra data transfers through the shared bus, and the extra memory accesses) accounts for 3.4\% of the overall energy consumption. Compared with ISAAC, there is a 13.5\% area overhead due to the estimation logics, the LUT, and the extra tiles occupied. However, due to the improvement of throughput, the area efficiency is improved by 2.5 times. For the 8-bit implementation of CifarQuick, 39.1\% of computations are reduced. Thus the energy efficiency and throughput are improved by 1.4 times and 1.6 times, respectively, at a cost of 0.15\% accuracy loss and 14.8\% area overhead. Similar results have been observed for LeNet-5. The improvements in energy efficiency and throughput are 3.0 times and 4.5 times for the 16-bit implementation. For the 8-bit implementation, the corresponding improvements are 1.6 and 1.9 times, respectively. Around 12\% area overhead is induced for the implementations of LeNet-5.

\section{Conclusions}

In this paper, a RRAM-based CNN accelerator is proposed to reduce the computations during the inference. The computations are reduced by exploiting the output sparsity and the adaptive approximation based on the runtime estimation on the maximum and minimum values of the output activation. It is implemented under different quantization schemes, and the corresponding energy efficiency and throughput are significantly improved.


\bibliographystyle{abbrv}

\end{document}